\documentclass[journal]{IEEEtran}
\usepackage{fancyhdr}
\usepackage{amsmath}
\usepackage{subfigure}
\usepackage{soul}
\usepackage{extarrows}
\usepackage{color}
\usepackage{mathrsfs}
\usepackage{graphicx}
\usepackage{booktabs}
\usepackage{algorithm} 
\usepackage{algorithmic} 
\begin{document}
\bibliographystyle{IEEEtran}
\title{Smart Pilot Assignment for Massive MIMO\\
}
\author{\IEEEauthorblockN{Xudong Zhu, Zhaocheng Wang, Linglong Dai, and Chen Qian
\vspace*{-8mm}
}

\thanks{X. Zhu, Z. Wang, L. Dai, and C. Qian are with Tsinghua National Laboratory for Information Science and Technology (TNList), Department of Electronic Engineering, Tsinghua University, Beijing 100084 (E-mails: zhuxd12@mails.tsinghua.edu.cn, zcwang@tsinghua.edu.cn, daill@tsinghua.edu.cn, qianc10@mails.tsinghua.edu.cn).}

\thanks{This work was supported by National High Technology Research and Development Program of China (Grant No. 2014AA01A704), National Key Basic Research Program of China (Grant No. 2013CB329203), Beijing Natural Science Foundation (Grant No. 4142027), National Nature Science Foundation of China (Grant No. 61271266) and the Foundation of Shenzhen government.}
}
\maketitle
\begin{abstract}
A massive multiple-input multiple-output (MIMO) system, which utilizes a large number of antennas at the base station (BS) to serve multiple users, suffers from pilot contamination due to inter-cell interference.
A smart pilot assignment (SPA) scheme is proposed in this letter to improve the performance of users with severe pilot contamination.
Specifically, by exploiting the large-scale characteristics of fading channels, the BS firstly measures the inter-cell interference of each pilot sequence caused by the users with the same pilot sequence in other adjacent cells.
Then, in contrast to the conventional schemes which assign the pilot sequences to the users randomly, the proposed SPA method assigns the pilot sequence with the smallest inter-cell interference to the user having the worst channel quality in a sequential way to improve its performance.
Simulation results verify the performance gain of the proposed scheme in typical massive MIMO systems.
\end{abstract}

\IEEEpeerreviewmaketitle

\section{Introduction}
Massive multiple-input multiple-output (MIMO) has been recently investigated to meet the exponential increase of mobile traffic in wireless systems \cite{noncooperative}, \cite{scalingupMIMO}, whereby a base station (BS) equipped with a large number of antennas serves multiple users simultaneously.
Asymptotic analysis based on random matrix theory \cite{scalingupMIMO} demonstrates that the intra-cell interference and the uncorrelated noise can be effectively eliminated when the number of BS antennas goes to infinity.
However, pilot contamination caused by the inter-cell interference originated from the reuse of the same pilot group in adjacent cells, does not vanish as the increase of BS antennas, and it becomes the performance bottleneck of massive MIMO systems \cite{noncooperative}, \cite{scalingupMIMO}.

The issue of pilot contamination has been widely studied in the literature \cite{timeshift1}-\cite{blind1}.
The time-shifted pilot scheme is an effective solution by using asynchronous transmission among adjacent cells \cite{timeshift1}, but leads to the mutual interferences between data and pilot.
A greedy pilot assignment algorithm \cite{JS} can mitigate the pilot contamination by exploiting the statistical channel covariance information, but suffers from high computational complexity of iteratively minimizing the mean degree of spatial orthogonality.
Pilot contamination precoding \cite{PCP1} can mitigate the inter-cell interference by multi-cell joint processing, but suffers from spectral efficiency loss due to high overhead required by information exchange.
In addition, a blind method based on subspace partitioning \cite{blind1} is able to reduce the inter-cell interference when the channel vectors from different users are orthogonal.
All those solutions assign the available pilot sequences to different users randomly without considering their different channel qualities, and also ignore the fact that the severity of pilot contamination varies among different pilot sequences.

In this letter, a smart pilot assignment (SPA) scheme is proposed to enhance the performance of users with severe pilot contamination.
We consider the pilot assignment problem for a target cell, which is surrounded by other adjacent cells.
Unlike the conventional schemes which assign the available pilot sequences to the users in a random way, our proposed SPA scheme aims to maximize the minimum uplink signal-to-interference-plus-noise-ratio (SINR) of all users in the target cell.
Specifically, by exploiting the large-scale characteristics of fading channels, the BS firstly measures the inter-cell interference of each pilot sequence caused by the users with the same pilot sequence in other adjacent cells.
After that, the channel qualities from different users in the target cell to the BS can be detected, which usually differ from one user to another.
The proposed SPA method assigns the pilot sequence with the smallest inter-cell interference to the user having the worst channel quality in a sequential way until all users have been assigned by their corresponding pilot sequences.
Simulation results verify the effectiveness and performance gain of the proposed SPA scheme in typical massive MIMO systems.

\section{System Model}

\begin{figure}
\vspace{0.1cm}
\center{\includegraphics[angle=90, width=0.45\textwidth]{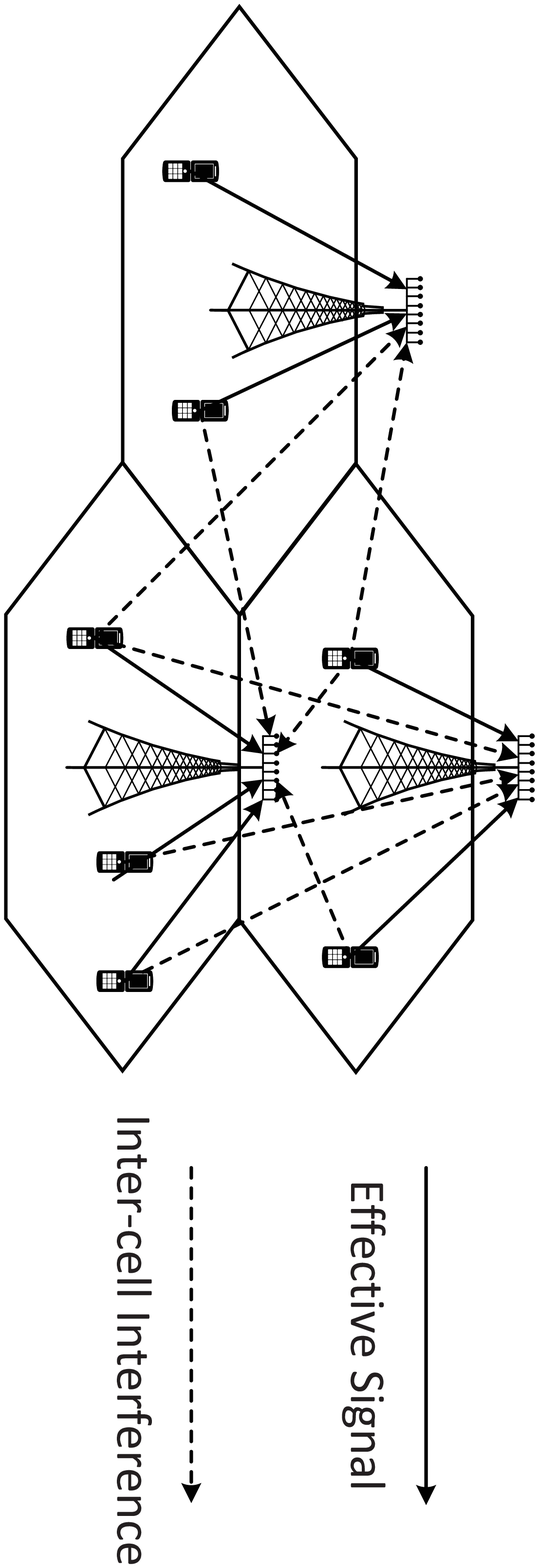}}
\vspace{-0.2cm}
\caption{A typical multi-cell multi-user massive MIMO systems.}
\label{systemmodel}
\vspace{-0.5cm}
\end{figure}

As shown in Fig. \ref{systemmodel}, we consider a multi-cell multi-user massive MIMO system composed of $L$ hexagonal cells, and each cell is consisted of a BS with $M$ antennas and $K$ ($K\ll M$) single-antenna users \cite{noncooperative}, \cite{scalingupMIMO}.
The channel vector $\mathbf{h}_{ijk}\in\mathcal{C}^{M\times 1}$ from the $k$-th user in the $j$-th cell to the BS in the $i$-th cell can be modeled as
\begin{equation}
\mathbf{h}_{ijk} = \mathbf{g}_{ijk}\sqrt{\beta_{ijk}},
\end{equation}
where $\beta_{ijk}$ denotes the large-scale fading coefficients which change slowly and can be easily tracked \cite{JS}-\cite{blind1}, and $\mathbf{g}_{ijk}\sim\mathcal{CN}(\mathbf{0},\mathbf{I}_M)$ denotes the small-scale fading vectors.
By considering the typical protocol of time-division duplexing (TDD) in massive MIMO systems, we adopt the widely used time block fading model, whereby the channel vector $\mathbf{h}_{ijk}$ remains constant during the coherence interval \cite{noncooperative}, \cite{scalingupMIMO}.

Assuming the pilot sequences $\mathbf{\Phi}=[\boldsymbol{\phi}_1, \boldsymbol{\phi}_2, \cdots, \boldsymbol{\phi}_K]^T\in\mathcal{C}^{K\times\tau}$ with length of $\tau$ used in one cell are orthogonal, i.e., $\mathbf{\Phi}\mathbf{\Phi}^H=\mathbf{I}_K$, and the same pilot group is reused in other cells due to the limited pilot resource \cite{noncooperative}.
The conventional pilot assignment methods usually assign the pilot sequence $\boldsymbol{\phi}_k$ to the $k$-th user without considering the different channel qualities among users \cite{noncooperative}, \cite{scalingupMIMO}.
Thus, the received pilot sequence $\mathbf{Y}_i^{\text{p}}\in\mathcal{C}^{M\times \tau}$ at the BS in the $i$-th cell can be represented as
\begin{equation}
\label{receivedpilotsequence}
\mathbf{Y}_i^{\text{p}} = \sqrt{\rho_{\text{p}}}\sum_{j=1}^{L}\sum_{k=1}^{K}\mathbf{h}_{ijk}\boldsymbol{\phi}_k^T+\mathbf{N}_i^{\text{p}},
\end{equation}
where $\rho_{\text{p}}$ denotes the pilot transmission power, and $\mathbf{N}_i^{\text{p}}\in\mathcal{C}^{M\times\tau}$ denotes the additive Gaussian white noise (AWGN) matrix with entries being independent and identically distributed (i.i.d.) Gaussian random variables with zero-mean and variance $\sigma_{N}^2$.
Similarly, the received user data $\mathbf{y}_i^{\text{u}}\in\mathcal{C}^{M\times 1}$ at the BS in the $i$-th cell can be represented as
\begin{equation}
\label{receivedsignal}
\mathbf{y}_i^{\text{u}} = \sqrt{\rho_\text{u}}\sum_{j=1}^{L}\sum_{k=1}^{K} \mathbf{h}_{ijk}x_{jk}^{\text{u}}+\mathbf{n}_i^{\text{u}},
\end{equation}
where $x_{jk}^{\text{u}}$ denotes the symbol from the $k$-th user in the $j$-th cell with $\text{E}\{|x_{jk}^{\text{u}}|^2\}=1$, $\rho_{\text{u}}$ denotes the uplink data transmission power, and $\mathbf{n}_i^{\text{u}}\in\mathcal{C}^{M\times 1}$ denotes the AWGN vector with $E\{\mathbf{n}_i^{\text{u}}(\mathbf{n}_i)^H\}=\sigma_n^2\mathbf{I}_M$.
By correlating the received pilot sequence $\mathbf{Y}_i^{\text{p}}$ with pilot sequence $\boldsymbol{\phi}_k$, the channel estimate of the $k$-th user in the $i$-th cell can be represented as
\begin{equation}
\label{ce}
\hat{\mathbf{h}}_{iik}=\frac{1}{\sqrt{\rho_{\text{p}}}}\mathbf{Y}_i^{\text{p}}\boldsymbol{\phi}_k^H
=\sum_{j=1}^{L}\mathbf{h}_{ijk}+\mathbf{v}_{ik},
\end{equation}
where $\mathbf{v}_{ik}=\frac{1}{\sqrt{\rho_{\text{p}}}}\mathbf{N}_i^{\text{p}}\boldsymbol{\phi}_k^H$ denotes the equivalent noise.
It is clear that the channel estimate of the $k$-th user in the $i$-th cell, $\hat{\mathbf{h}}_{iik}$, is a linear combination of the channels $\mathbf{h}_{ijk}$, $j=1,2,\cdots,L$, of the users with the same pilot sequence in all cells, which is referred to as pilot contamination \cite{noncooperative}, \cite{scalingupMIMO}.

By adopting the matched-filter (MF) detector based on the channel estimate result $\hat{\mathbf{h}}_{iik}$, the detected symbol for the $k$-th user in the $i$-th cell can be represented as
\begin{eqnarray}
\label{uplinkdatadetection}
\hat{x}_{ik}^{\text{u}}&=&\hat{\mathbf{h}}_{iik}^H\mathbf{y}_i^{\text{u}}\nonumber\\
&=&(\sum_{j=1}^L\mathbf{h}_{ijk}+\mathbf{v}_{ik})^H(\sqrt{\rho_{\text{u}}}\sum_{j=1}^{L}\sum_{k^{'}=1}^{K} \mathbf{h}_{ijk^{'}}x_{jk^{'}}^{\text{u}}+\mathbf{n}_i^{\text{u}})\nonumber\\
&=& \sqrt{\rho_{\text{u}}}(\mathbf{h}_{iik}^H\mathbf{h}_{iik} x_{ik}^{\text{u}}+\sum_{j\neq i}\mathbf{h}_{ijk}^H\mathbf{h}_{ijk} x_{jk}^{\text{u}})+\varepsilon_{ik}^{\text{u}},
\end{eqnarray}
where $\varepsilon_{ik}^{\text{u}}$ denotes the intra-cell interference and uncorrelated noise, which can be significantly reduced by increasing the number of BS antennas \cite{scalingupMIMO}.
Then, the uplink SINR of the $k$-th user in the $i$-th cell can be calculated as
\begin{equation}
\label{SINRUplink}
\text{SINR}_{ik}^{\text{u}} = \frac{|\mathbf{h}_{iik}^H\mathbf{h}_{iik}|^2}{\sum_{j\neq i}|\mathbf{h}_{ijk}^H \mathbf{h}_{ijk}|^2 +\frac{|\varepsilon_{ik}^{\text{u}}|^2}{\rho_{\text{u}}}}
\xlongrightarrow{M\rightarrow\infty}\frac{\beta_{iik}^2}{\sum_{j\neq i}\beta_{ijk}^2}.
\end{equation}

Thus, the corresponding average uplink capacity of this user can be calculated as $\text{C}_{ik}^{\text{u}}=\text{E}\{\log_2(1+\text{SINR}_{ik}^{\text{u}})\}$.
It is clear that thermal noise and small-scale fading effects could be averaged out as $M$ grows to infinity.
However, the average uplink capacity is limited by the pilot contamination and cannot be improved by the increase of either $\rho_{\text{u}}$ or $\rho_{\text{p}}$.

\section{Proposed Scheme}
In this section, the pilot assignment for a target cell is firstly formulated as an optimization problem.
Then, the SPA scheme is proposed to approach the optimization solution in a greedy way.
The performance analysis is also provided to verify the effectiveness of the proposed SPA scheme.

\subsection{Problem Formulation}
We consider the pilot assignment for a specific cell, i.e., the $i$-th cell as the target cell, and the pilot assignment for other cells are independently managed by their corresponding BSs.
For this target cell, the number of different kinds of pilot assignments between $K$ users $[U_{1},U_{2},\cdots,U_{K}]$ and $K$ pilot sequences $[\boldsymbol{\phi}_1, \boldsymbol{\phi}_2, \cdots, \boldsymbol{\phi}_K]$ is huge, i.e., $P(K,K)=K!$.
In the conventional pilot assignment schemes \cite{noncooperative}, \cite{scalingupMIMO}, the pilot sequence $\boldsymbol{\phi}_k$ is assigned to the $k$-th user $U_k$ randomly.
Based on the possible cooperation among cells, such as coordinated multiple points (CoMP) in LTE-A systems, a specific BS is able to acquire the pilot assignment knowledge of other cells.
Without loss of generality, for other cells around the target cell, we assume that the random pilot assignment is used.

In massive MIMO systems, since the poor users with severe pilot contamination are the performance bottleneck \cite{noncooperative}, \cite{scalingupMIMO}, we aim to maximize the minimum uplink SINR of all $K$ users in the target cell, which can be formulated as the following optimization problem
\begin{equation}
\label{op1}
\mathcal{P}:\max_{\{\mathcal{F}_s\}}\hspace{0.1cm} \min_{\forall k}
\frac{|\mathbf{h}_{iif_s^k}^H\mathbf{h}_{iif_s^k}|^2}{\sum_{j\neq i}|\mathbf{h}_{ijk}^H \mathbf{h}_{ijk}|^2 +\frac{|\varepsilon_{if_s^k}^{\text{u}}|^2}{\rho_{\text{u}}}},
\end{equation}
where $\{\mathcal{F}_s: s=1,\cdots,K!\}$ denotes all possible $K!$ kinds of pilot assignments, e.g., $\mathcal{F}_s=[f_s^1,f_s^2,\cdots,f_s^K]$ denotes the $s$-th assignment, and assuming that the pilot sequence $\boldsymbol{\phi}_k$ is assigned to the $k$-th user $U_k$ in all other cells.

However, it seems impossible to solve this optimization problem $\mathcal{P}$ due to the fact that we cannot obtain accurate channel estimate under pilot contamination as shown in (\ref{ce}).
Fortunately, the limit of the uplink SINR can be represented by the large-scale fading coefficients $\beta_{ijk}$ as shown in (\ref{SINRUplink}). 
As shown in \cite{JS}-\cite{blind1}, the large-scale fading coefficients $\beta_{ijk}$ change slowly and can be easily tracked by the BSs.
Thus, the optimization problem $\mathcal{P}$ can be approached by
\begin{equation}
\label{op2}
\mathcal{P} \xlongrightarrow{M\rightarrow\infty}\mathcal{P}':\max_{\{\mathcal{F}_s\}}\hspace{0.1cm} \min_{\forall k}
\frac{\beta_{iif_s^k}^2}{\sum_{j\neq i}\beta_{ijk}^2}.
\end{equation}
The proposed SPA scheme aims to solve this optimization problem $\mathcal{P}'$ in a greedy way, which will be addressed in detail in the next subsection.

\subsection{Pilot Assignment}
The most direct way to solve the optimization problem $\mathcal{P}'$ is the exhaustive search, which tries all possible assignments and chooses the best one.
However, the number of all pilot assignments is as huge as $K!$, which leads to high computational complexity.
In this letter, the optimization problem $\mathcal{P}'$ is solved in a greedy way with low complexity.

For the target cell, we define a series of parameters $\{\alpha_k\}_{k=1}^{K}$ to quantify the channel quality of $K$ users as
\begin{equation}
\alpha_k=\beta_{iik}^2,\hspace{0.2cm} k=1,2,\cdots,K.
\end{equation}
For the $K$ pilot sequences $[\boldsymbol{\phi}_1,\boldsymbol{\phi}_2,\cdots,\boldsymbol{\phi}_K]$, we define another series of parameters $\{\gamma_k\}_{k=1}^{K}$ to quantify the inter-cell interference of each pilot sequence caused by the users with same pilot sequence in other adjacent cells as
\begin{equation}
\gamma_{k}=\sum_{j\neq i}\beta_{ijk}^2, \hspace{0.2cm} k=1,2,\cdots,K,
\end{equation}
which varies among $K$ pilot sequences.

For a specific assignment $\mathcal{F}_s=[f_s^1,f_s^2,\cdots,f_s^K]$, the pilot sequence $\boldsymbol{\phi}_k$ is assigned to the user $U_{f_s^k}$, whereby the limit of uplink SINR of the user $U_{f_s^k}$, i.e., $\text{SINR}_{ik}^{\text{u}}\rightarrow\alpha_{f_s^k}/\gamma_{k}$, is decided by two aspects: 1) the channel quality $\alpha_{f_s^k}$ of the user $U_{f_s^k}$; 2) the inter-cell interference $\gamma_{k}$ caused by the users with the same pilot sequence $\boldsymbol{\phi}_k$ in adjacent cells.
In order to maximize the minimum uplink SINR of all users in the target cell, we need to avoid the pilot sequence with great inter-cell interference assigned to the user having bad channel quality, which leads to relatively low uplink SINR.

Based on this motivation, the SPA scheme is proposed, which assigns the pilot sequence with the smallest inter-cell interference to the user having the worst channel quality in a sequential way.
Mathematically, as shown in Fig. \ref{SPA}, we first sort the $K$ pilot sequences according to the severity of inter-cell interference in descending order, i.e.,
\begin{equation}
\label{f_p}
\mathcal{F}_p: [\boldsymbol{\phi}_{f_p^1},\boldsymbol{\phi}_{f_p^2},\cdots,\boldsymbol{\phi}_{f_p^K}],
\end{equation}
whereby this permutation $\mathcal{F}_p$ of the $K$ pilot sequences satisfies
$\gamma_{f_p^1}\geq\gamma_{f_p^2}\geq\cdots\geq\gamma_{f_p^K}>0$.
Similarly, we sort the $K$ users according to their channel qualities in descending order, i.e.,
\begin{equation}
\label{f_q}
\mathcal{F}_q: [U_{f_q^1},U_{f_q^2},\cdots,U_{f_q^K}],
\end{equation}
whereby this permutation $\mathcal{F}_q$ of the $K$ users satisfies
$\alpha_{f_q^1}\geq\alpha_{f_q^2}\geq\cdots\geq\alpha_{f_q^K}>0$.
Then the pilot sequence $\boldsymbol{\phi}_{f_p^k}$ will be assigned to the corresponding user $U_{f_q^k}$.
Apparently, the computational complexity of the proposed SPA scheme comes from the sorting process, which is only $\mathcal{O}(K \log K)$ and negligible compared with $\mathcal{O}(K!)$ required by exhaustive search.

In order to apply the proposed greedy solution to the whole system with $L$ cells, a sequential iterative scheme can be utilized.
Specifically, all $L$ cells solve their own optimization problems in a sequential way, and then this sequential procedure is carried out iteratively until the convergence is achieved.

\subsection{Performance Analysis}

\begin{figure}
\vspace{-0.1cm}
\center{\includegraphics[angle=90,width=0.47\textwidth]{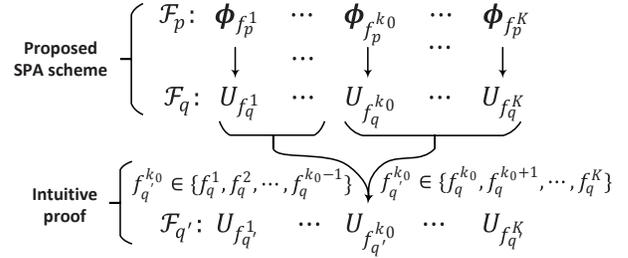}}
\vspace{-0.2cm}
\caption{Intuitive description and proof of the proposed SPA scheme.}
\label{SPA}
\vspace{-0.2cm}
\end{figure}

In this subsection, we prove that the pilot assignment greedily generated by the proposed SPA scheme is one of the solutions to the optimization problem $\mathcal{P}'$.

Without loss of generality, we fix the permutation of $K$ pilot sequences as $\mathcal{F}_p$ addressed in (\ref{f_p}), and assume that there is another permutation $\mathcal{F}_{q'}$ of $K$ users, which has larger minimum uplink SINR of $K$ users than that of $\mathcal{F}_{q}$, i.e.,
\begin{equation}
\label{assumption}
\mu_{q'}=\min_{\forall k}
\frac{\alpha_{f_{q'}^k}}{\gamma_{f_{p}^k}}
>\mu_q=\min_{\forall k}
\frac{\alpha_{f_q^k}}{\gamma_{f_{p}^k}}.
\end{equation}

Now we prove that such permutation $\mathcal{F}_{q'}$ does not exist and the intuitive proof is shown in Fig. \ref{SPA}.
More specifically, we firstly assume that the user $U_{f_{q}^{k_0}}$ has the minimum uplink SINR in the permutation $\mathcal{F}_{q}$, i.e., $\mu_q=\alpha_{f_q^{k_0}}/\gamma_{f_{p}^{k_0}}$.
Then, according to the permutation $\mathcal{F}_{q'}$, the pilot sequence $\boldsymbol{\phi}_{f_p^{k_0}}$ will be assigned to the user $U_{f_{q'}^{k_0}}$.
Comparing $\alpha_{f_{q'}^{k_0}}$ with $\alpha_{f_q^{k_0}}$, there will be two results:
\begin{enumerate}
  \item $\alpha_{f_{q'}^{k_0}}\leq\alpha_{f_{q}^{k_0}}$: Due to the fact that the permutation of the pilot sequences $\mathcal{F}_p$ is fixed, we have
      \begin{equation}
          \mu_{q'} \leq \frac{\alpha_{f_{q'}^{k_0}}}{\gamma_{f_{p}^{k_0}}}
          \leq \frac{\alpha_{f_{q}^{k_0}}}{\gamma_{f_{p}^{k_0}}} = \mu_q,
      \end{equation}
      which contradicts with the assumption in (\ref{assumption}).
  \item $\alpha_{f_{q'}^{k_0}}>\alpha_{f_{q}^{k_0}}$: Due to the sorting process in (\ref{f_q}), it is clear that
      \begin{equation}
            f_{q'}^{k_0}\in\{f_q^1,f_q^2,\cdots,f_q^{k_0-1}\}.
      \end{equation}
      Then, we consider the pilot assignment of the pilot sequences $\{\boldsymbol{\phi}_{f_p^1},\boldsymbol{\phi}_{f_p^2},\cdots,\boldsymbol{\phi}_{f_p^{k_0-1}}\}$.
      Since one user chosen from $\{U_{f_q^1},U_{f_q^2},\cdots,U_{f_q^{k_0-1}}\}$ has been assigned with the pilot sequence $\boldsymbol{\phi}_{f_p^{k_0}}$, there must be an user $U_{f_q^{k_1}}$ ($k_1\geq k_0$) assigned with a pilot sequence $\boldsymbol{\phi}_{f_p^{k_2}}$ ($k_2<k_0$).
      As a result, we have
      \begin{equation}
            \mu_{q'}\leq \frac{\alpha_{f_q^{k_1}}}{\gamma_{f_p^{k_2}}} \leq \frac{\alpha_{f_q^{k_0}}}{\gamma_{f_p^{k_2}}}
            \leq \frac{\alpha_{f_q^{k_0}}}{\gamma_{f_p^{k_0}}} =\mu_q,
      \end{equation}
      which contradicts with the assumption in (\ref{assumption}).
\end{enumerate}
It is concluded that the permutation $\mathcal{F}_q$ generated by the proposed SPA scheme is one of the solutions to the optimization problem $\mathcal{P}'$.
Note that there may be more than one optimal solution, for example, there are two users with the same channel quality, i.e., $\alpha_{k_1}=\alpha_{k_2}, k_1\neq k_2$. However, the exchange of the pilot sequences assigned to these two users makes no difference on the minimum uplink SINR.

It should be pointed out that the solution to the optimization problem $\mathcal{P}'$ is also the solution to the optimization problem $\mathcal{P}$ when the number of antennas at the BS grows to infinity, i.e., $M\rightarrow\infty$.
With a finite number of BS antennas in practical systems, the solution to $\mathcal{P}'$ can approach the solution to $\mathcal{P}$, which will be verified by simulation results.

\section{Numerical Results}

\begin{figure*}[!t]
\centering
\vspace{-0.2cm}
\subfigure[] {\includegraphics[height=1.9in,width=2.3in]{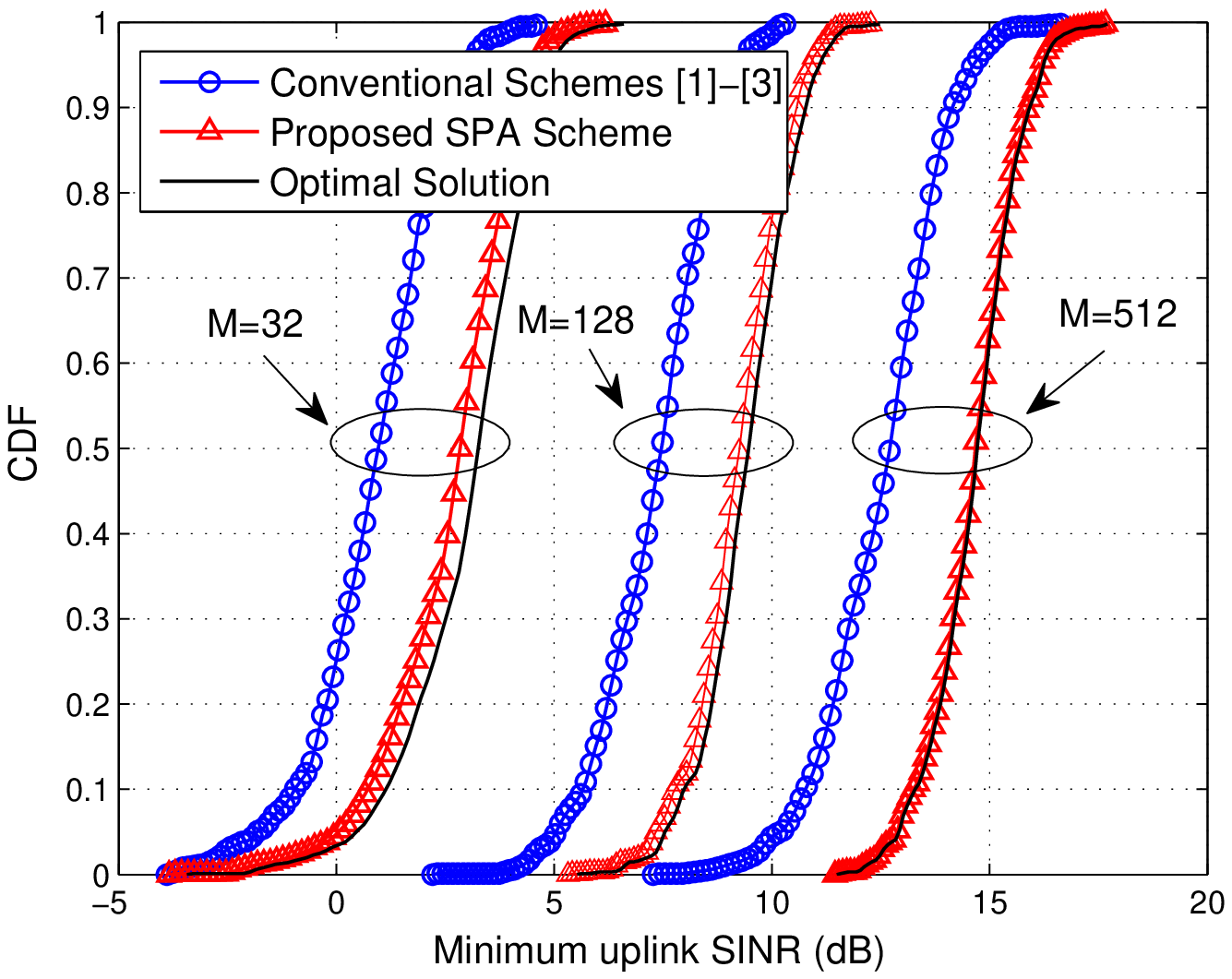}}
\subfigure[] {\includegraphics[height=1.9in,width=2.3in]{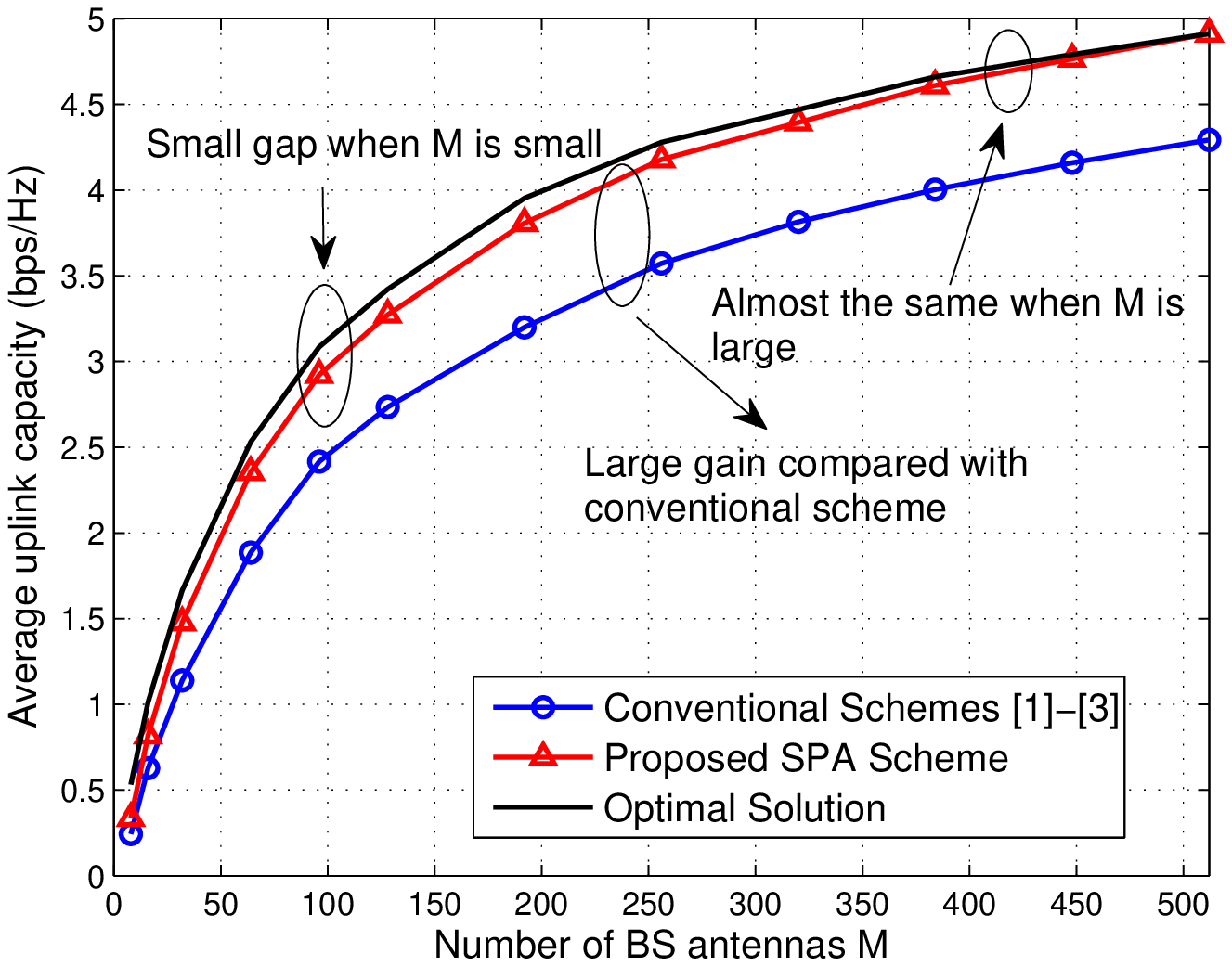}}
\subfigure[] {\includegraphics[height=1.9in,width=2.3in]{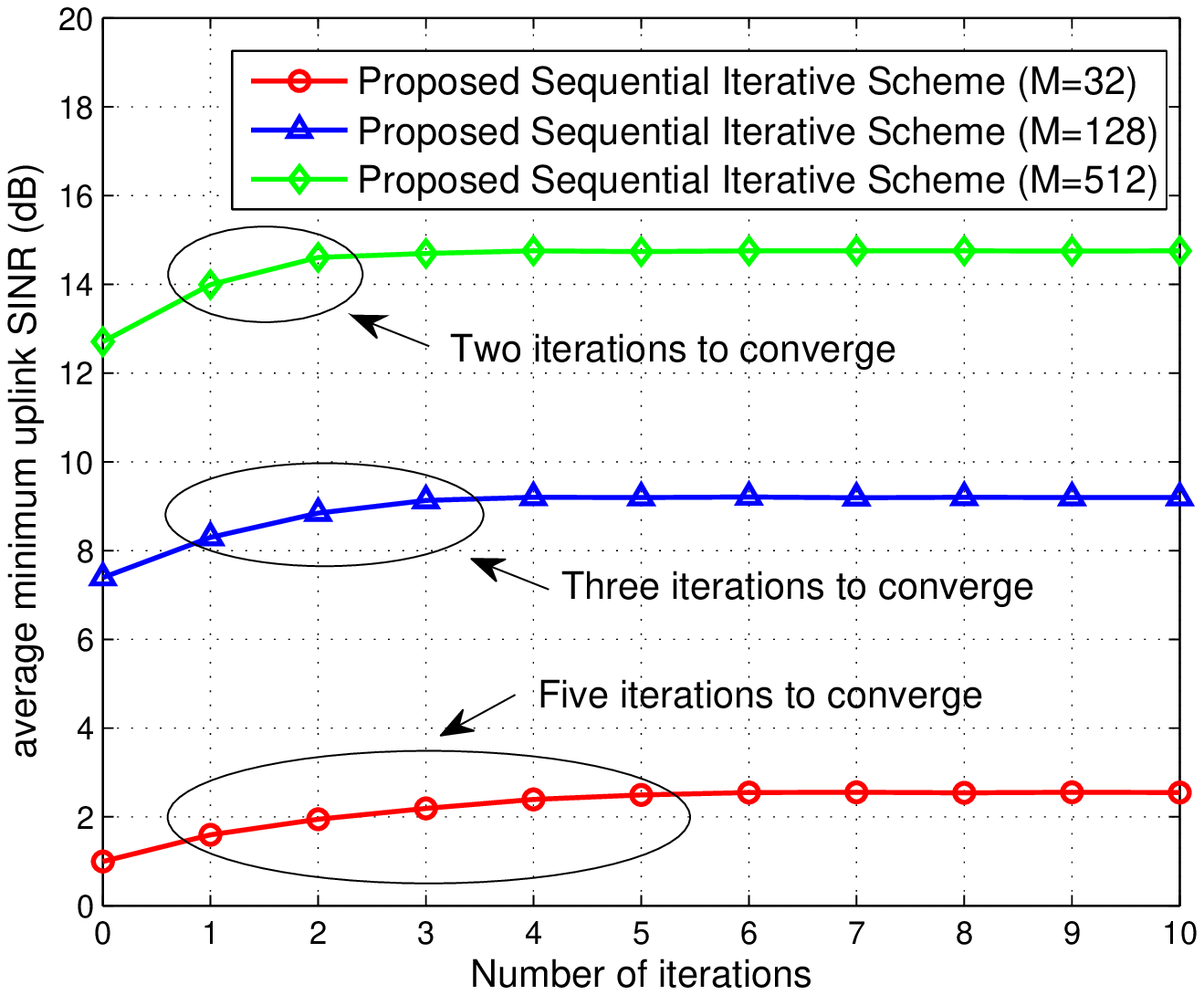}}
\vspace{-0.2cm}
\caption{Simulation results: (a) The CDF of the minimum uplink SINR among all $K$ users in the target cell; (b) The average uplink capacity of the user with the minimum uplink SINR among in the target cell;
(c) The convergence of the average minimum uplink SINR in each cell.}
\label{FigSimulation}
\vspace{-0.4cm}
\end{figure*}

In this section, we investigate the performance of the proposed SPA scheme through Monte-Carlo simulations.
A typical hexagonal cellular network with $L$ cells is considered, where each cell has $K$ users with single-antenna and a BS with $M$ antennas \cite{noncooperative}, \cite{scalingupMIMO}.
The center cell surrounded by other cells is considered as the target cell.
The system parameters are summarized in Table I.
As addressed in \cite{scalingupMIMO}, the large-scale fading coefficient $\beta_{ijk}$ can be modeled as
\begin{equation}
\beta_{ijk}={z_{ijk}}/{(r_{ijk}/R)^{\alpha}},
\end{equation}
where $z_{ijk}$ represents the shadow fading and possesses a log-normal distribution (i.e., $10\log_{10}(z_{ijk})$ is Gaussian distributed with zero mean and the standard deviation of $\sigma_{\text{shadow}}$), $r_{ijk}$ is the distance between the $k$-th user in the $j$-th cell and the BS in the $i$-th cell, and $R$ is the cell radius.

\begin{table}[htbp]
\centering
\label{parameters}
\caption{Simulation Parameters}
\begin{tabular}{l|l}
  \hline\hline
  Number of cells $L$ & 7 \\\hline
  Number of BS antennas $M$ & $8\leq M\leq 512$ \\\hline
  Number of users in each cell $K$ & $K=8$ \\\hline
  Cell radius $R$ & 500 m\\\hline
  Cell edge SNR & 20 dB \\\hline
  Average transmit power at users $\rho_{\text{p}},\rho_{\text{u}}$ & 0 dBm \\\hline
  Path loss exponent $\alpha$ & 3 \\\hline
  Log normal shadowing fading $\sigma_{\text{shadow}}$ & 8 dB \\\hline
  \hline
\end{tabular}
\end{table}

Fig. \ref{FigSimulation} (a) plots the cumulative distribution function (CDF) curve of the minimum uplink SINR among all $K$ users in the target cell.
The conventional schemes assign the pilot sequence $\boldsymbol{\phi}_k$ to the user $U_k$ \cite{noncooperative}, \cite{scalingupMIMO}, and the optimal solution is the solution to the optimization problem $\mathcal{P}$ obtained by exhaustive search.
When $M=32$ is considered, it is evident that the proposed SPA scheme outperforms the conventional schemes by about 2 dB, and the optimal solution is the best with another gain about 0.5 dB.
When the number of BS antennas increases, i.e., $M=512$ is considered, we can find that the performance of the proposed SPA scheme is almost the same as that of the optimal solution.

Fig. \ref{FigSimulation} (b) shows the average uplink capacity of the user with the minimum uplink SINR among all $K$ users in the target cell.
When the typical parameter $M=128$ is considered, the average uplink capacity of the user with the minimum uplink SINR in the proposed SPA scheme is larger than that of the conventional schemes by about 0.6 bps/Hz.
Moreover, when $M$ grows large, the performance of the proposed SPA scheme approaches that of the optimal solution.

Fig. \ref{FigSimulation} (c) shows the convergence of the average minimum uplink SINR in each cell when the sequential iterative scheme is applied to the whole system.
When $M=32$ is considered, it takes about 5 iterations to converge for all $L$ cells, and the required number of iterations becomes smaller if the number of BS antennas increases, e.g., $M=128$ or $M=512$ as illustrated in Fig. \ref{FigSimulation} (c).

\section{Conclusions}
In this letter, we have proposed a smart pilot assignment scheme to improve the minimum uplink SINR of all users within the target cell in massive MIMO systems.
By exploiting the large-scale characteristics of fading channels, the proposed SPA scheme assigns the pilot sequence with the smallest inter-cell interference to the user having the worst channel quality in a sequential way.
Theoretical analysis proves that the pilot assignment generated by the SPA scheme is the solution to the simplified optimization problem $\mathcal{P}'$, which can approach the original optimization problem $\mathcal{P}$ when $M$ grows to infinity.
Simulation results demonstrate that for the typical configuration of 128 BS antennas in multi-user massive MIMO systems, the proposed SPA scheme is able to improve the minimum uplink SINR of users by about 2 dB.

\end{document}